\documentclass[prl,showpacs,amsfonts,amssymb,amsmath]{revtex4}
\usepackage{times}
\usepackage{graphicx}
\newcommand{\lambdav}{\mbox{\boldmath$\lambda$}}
\newcommand{\nablav}{\mbox{\boldmath$\nabla$}}

\begin{document}

\title{Force on a moving point impurity due to quantum fluctuations in a Bose-Einstein condensate}

\author{D. C. Roberts} \affiliation{Laboratoire de
  Physique Statistique de l'Ecole Normale Sup\'erieure, Paris, France} 
\begin{abstract}
An analytic expression is derived for a force on a weak point impurity arising from the scattering of quantum fluctuations in a slow-moving, weakly interacting, three-dimensional Bose-Einstein condensate at zero temperature.  In an infinitely extended geometry, this force is shown to exist at any arbitrarily small flow velocity below Landau's critical velocity.  Furthermore, this force is shown to be directly proportional to the flow speed. 
\end{abstract}

\maketitle

The exotic properties of superfluid flow, especially the existence of a critical velocity below which the superfluid flows without dissipation \cite{landau}, continue to be the subject of active investigation after more than half a century.   A recent letter \cite{roberts} argued that the scattering of zero-temperature quantum fluctuations could contribute to a force that is present at all flow velocities, adding a new facet to this complex subject.   Specifically, the effective critical velocity, defined by the minimum flow velocity at which an immersed object feels a force from the flow, was shown to be zero for a one-dimensional object in an infinitely extended superfluid modeled by a three-dimensional weakly-interacting Bose-Einstein condensate.   However, the idealized 1-d potential used in this example has many peculiarities - such as the possibility of bound states, the presence of scattered waves at infinity,  the force being negative and the changing sign of the effective mass - that potentially cloud the main idea that a drag force can be caused by quantum fluctuations at zero temperature.   In this letter, we describe the calculation of a more experimentally realizable and theoretically transparent potential, namely that of a weak three-dimensional repulsive point impurity, which avoids the complications mentioned above.  This calculation would be directly applicable to experiments involving trapped dilute BECs where untrapped atoms act as impurities \cite{impurity}.  We derive a simple analytical result that shows that the force on a stationary weak point impurity in the flow of an infinitely extended weakly interacting condensate is proportional to the flow speed and exists at any any arbitrarily small flow speed, and discuss how in a finite geometry this force may eventually cancel out due to backscattering effects.  

  We consider a three-dimensional, weakly interacting condensate of density $n_0$, characterized by a delta-function interparticle contact pseudopotential whose coupling constant $g$ is determined by the 2-particle positive condensate atom-atom scattering length $a$ and the mass $m$ of the atoms, i.e. $g=4 \pi \hbar^2 a/m$, flowing at a speed $c$ at infinity relative to a fixed weak impurity.  Throughout this analysis, we assume the flow speed $c$ to be less than the speed of sound. Although temperature is not well-defined in the scattering problem detailed in this paper, we use ``$T=0$'' as a convenient description of the quantum state of the flowing condensate.  While the scenario of a fixed impurity in a moving condensate is equivalent to that of a moving impurity in a fixed condensate, differing only by a Galilean transformation, we perform the calculation in the fixed frame of the impurity where the Hamiltonian is time independent.  The potential of this fixed point impurity is described by $\eta \delta^{(3)}({\bf r}) $ where the coupling constant is given by $\eta=2 \pi \hbar^2 b/m$, $b$ being the scattering length that characterizes atom-impurity 2-body collisions; $\eta$ and $b$ are defined as positive.    We assume the condensate is dilute such that $\sqrt{n_0 a^3} \ll 1$ where $n_0$ is the condensate density and the condensate's interaction with the impurity is weak.

The general force arising from the flowing condensate on the fixed point impurity can be written in second quantized notation as
\begin{equation}
{\bf F}=-\int d^3{\bf r} \langle \hat \psi^\dag ({\bf r})[ \nablav  \eta \delta^{(3)}({\bf r}) ]\hat \psi ({\bf r})  \rangle_{T=0} 
\end{equation}
where $\hat \psi({\bf r}) $ and $\hat \psi^\dag({\bf r}) $ are field operators that describe the weakly interacting BEC flow and obey the standard boson commutation relations.   The leading order contribution to the force is zero at flow speeds lower than the speed of sound; this result can be calculated assuming that the field operators are approximated by a macroscopic classical field representing the condensate whose behavior is governed in this mean-field approximation by the Gross-Pitaevskii equation (GPE) \cite{astra}.  This absence of force below a critical velocity (in this case the speed of sound) verifies Landau's phenomenological picture, which does not explicitly include the quantum fluctuations.  Based on this picture, Landau argued that the moving ground state would decay only above a critical speed (and only) by the emission of quasiparticles; otherwise the flow would be dissipationless \cite{landau}.  Consequently, according to Landau's theory a stationary object in a superfluid flowing below this critical velocity would remain metastable and not experience any force.  

  However, if we go beyond this mean-field picture by including quantum fluctuations, this conclusion that there is no force at low flow velocities no longer holds, at least when considering an infinitely extended medium.   
We briefly review the general framework for calculating the force (described in detail in \cite{roberts}). 

 We approximate the Bose field operator that describes the flow by a large macroscopic classical field  plus a small quantum fluctuation operator (the small parameter in this case being given by the diluteness parameter $\sqrt{n_0 a^3}$ \cite{fetter}), i.e.
\begin{equation}
\hat \psi({\bf r})= \Psi({\bf r})+\hat \phi({\bf r}) 
\end{equation}
where $\Psi({\bf r}) $ is the macroscopic condensate that has been modified by the quantum fluctuations (see eq. \ref{GGPE}).  The force acting on an impurity can then be written as
\begin{equation}
{\bf F}={\bf F}_{fluc}+{\bf F}_{cond}=\eta \nablav[ \langle \hat \phi^\dag ({\bf r})  \hat \phi ({\bf r})  \rangle_{T=0}+|\Psi({\bf r}) |^2]_{r=0}
\end{equation}
where the term ${\bf F}_{fluc}$ is the contribution to the force coming directly from the fluctuations and the term ${\bf F}_{cond}$ is the contribution to the force coming from the condensate modified by the fluctuations.

We expand $\hat \phi$  in terms
of quasiparticle operators $\alpha_{{\bf k}}$ and $\alpha^\dag_{{\bf k}}$, which by
definition obey the standard boson commutation relations $[\hat
\alpha_{{\bf k}},\hat \alpha^\dag_{{\bf k}'}]=\delta_{{\bf k},{\bf k}'}$ where $\delta_{{\bf k},{\bf k}'}$ is the Kronecker delta function and $\bf k$ is the dimensionless momentum normalized by the healing length.
The quasiparticle
operators are weighted by the quasiparticle amplitudes
$u_{{\bf k}}({\bf r}) $ and $v_{{\bf k}}({\bf r}) $, i.e.
\begin{equation}
\label{qp} \hat \phi({\bf r}) =\sum_{{\bf k}'} \left( u_{{\bf k}}({\bf r})  \hat
\alpha_{{\bf k}} -v^*_{{\bf k}}({\bf r})  \hat \alpha_{{\bf k}}^\dag \right), 
\end{equation}
where the sum is taken over all excited states and excludes the condensate mode.  The normalization condition $\int d^3 {\bf r} \left( |u_{\bf k}({\bf r})|^2-|v_{\bf k}({\bf r})|^2 \right)=1$ is imposed to force that the quasiparticle operators to obey the usual bosonic commutation relations. In order for the quasiparticle operators to diagonalize the weakly interacting Hamiltonian, the quasiparticle amplitudes must satisfy the two sets of differential equations known as the Bogoliubov-de Gennes equations \cite{{fetter},{Bog}}:
\begin{equation}
\label{BdG1} 
\hat {\cal L} u_{{\bf k}}({\bf r})  -\Psi^2 v_{{\bf k}}({\bf r})  =
E_{{\bf k}} u_{{\bf k}}({\bf r}) 
\end{equation}
\begin{equation}
\label{BdG2} \hat {\cal L}^*  v_{{\bf k}}({\bf r}) - (\Psi^*)^2 u_{{\bf k}}({\bf r}) 
= -E_{\bf k} v_{{\bf k}}({\bf r}) ,
\end{equation}
where these equations are expressed in dimensionless variables with the length scale normalized by the healing length $\xi=(8 \pi n_0 a)^{-1/2}$\cite{dim}, $*$ denotes the complex conjugate, $\hat {\cal L} = \hat T + \bar{\eta} \delta^{(3)}({\bf r})  - \mu +2 |\Psi|^2$,  $\hat T = -\nabla^2+\sqrt{2} i {\bf q} \cdot \nablav +q^2/2$, the dimensionless speed is given by $q= c/c_s$, $c_s=\sqrt{n_0 g/m}$ is the speed of sound, $\mu=1+q^2/2$ is the chemical potential determined by imposing $\Psi^{(0)}(r)=1$ at $r=\infty$, and $\bar{\eta} =2b/ a n_0 \xi^3$ .  In this analysis, we assume that the point impurity is weak in the sense that $\bar{\eta} \ll 1$. The energy eigenvalue associated with momentum state $\bf{k}$ for the moving BEC flow is $E_k=\sqrt{2}qk_x+E_B$ where $E_B=k \sqrt{k^2+2}$ is the Bogoliubov dimensionless dispersion relation for a BEC at rest.  

The behavior of the condensate modified by the quantum fluctuations  is given by the generalized GPE \cite{Castin},
\begin{equation}
\label{GGPE}
 (\hat T+ \bar{\eta} \delta^{(3)}({\bf r}) -\mu) \Psi({\bf r}) +| \Psi({\bf r})  |^2 \Psi({\bf r}) + \sum_{{\bf k}'} \left[2 |v_{\bf k}({\bf r}) |^2\Psi({\bf r}) -  u_k({\bf r}) v^*_{\bf k}({\bf r}) \Psi({\bf r}) \right] -\chi({\bf r}) \Psi({\bf r}) =0.
\end{equation}
 The term proportional to $\sum_{\bf k'} u_{\bf k}({\bf r})v^*_{\bf k}({\bf r})$ is ultraviolet divergent because of the contact potential approximation and must be renormalized (see for example \cite{morgan}).  The last term, $\chi({\bf r})  \Psi({\bf r}) $, ensures the orthogonality between the excited modes and the condensate \cite{morgan}, where $\chi({\bf r}) =\sum_{{\bf k}'} c_{{\bf k}} v^*_{{\bf k}}({\bf r})$
where $c_{{\bf k}}=\int d^3r  | \Psi({\bf r})  |^2[\Psi^{*}({\bf r})  u_{{\bf k}}({\bf r}) + \Psi({\bf r})  v_{{\bf k}}({\bf r}) ]$.

The effective scattering problem for the quasiparticle amplitudes determined by the Bogoliubov-de Gennes equations (eqns. \ref{BdG1}, \ref{BdG2}) \cite{T0} can be solved perturbatively with $\bar{\eta}$ as the small parameter to give the total force on the impurity to leading order in $\bar{\eta}$:
\begin{equation}
F_{x}= 4 \sqrt{2}\pi^{-3/2} \bar{\eta}^2 p_0 \xi^2 \sqrt{n_0 a_{sc}^3} \int d^3 {\bf k} (f_{fluc}({\bf k})+ f_{cond}({\bf k}))
\end{equation}
where $f_{fluc}({\bf k})$ and $f_{cond}({\bf k})$ are the ${\bf k}$-dependent contributions of the force corresponding to ${\bf F}_{fluc}$ and ${\bf F}_{cond}$, respectively, and the zeroth order interaction pressure is given by $p_0=g n_0^2/2$.  As we have assumed the flow to be only in the $x$-direction it follows that the direction of the force must be parallel to the $x$-axis.  Henceforth, therefore, we drop the vector notation when referring to force.  Note that because the leading order contribution to the force is proportional to $\bar{\eta}^2$, it cannot be treated in a linear response formalism and a straightforward application of the fluctuation-dissipation theorem is not possible.
The {\bf k}-dependent contribution to the force arising directly from the quantum fluctuations is given by
\begin{equation}
f_{fluc}({\bf k})= i V_0 (k) \int d^{3} \lambdav  \lambda_x [\tilde V_1^*(\lambdav,{\bf k})- \tilde V_1(\lambdav,{\bf k})]
\end{equation}
and the {\bf k}-dependent contribution to the force from the modified condensate is given by
\begin{eqnarray}
&&f_{cond}({\bf k})= \int d^{3} \lambdav \frac{i \lambda_x}{\lambda^4+2 \lambda^2-\lambda_x^2 q^2} \nonumber\\ 
&& \left( \left \{\lambda^2[U_0(k)-4V_0(k)]+\sqrt{2}\lambda_x q U_0(k) \right \} [\tilde V^*_1(\lambdav,{\bf k})- \tilde V_1(\lambdav,{\bf k})]+V_0(k)[\lambda^2-\sqrt{2} \lambda_x q][\tilde U_1^*(\lambdav,{\bf k})- \tilde U_1(\lambdav,{\bf k})] \right)
\end{eqnarray}
where the tilde signifies the Fourier transform, i.e. $U_1({\bf r},{\bf k}) =\int d^3 \lambdav e^{i \lambda \cdot {\bf r}} \tilde U_1(\lambdav,{\bf k})$ and $V_1({\bf r},{\bf k}) =\int d^3 \lambdav e^{i \lambda \cdot  {\bf r}} \tilde V_1(\lambdav,{\bf k})$, and we have expanded to leading order in $\bar{\eta}$ such that $u_{\bf k}({\bf r}) =[U_0({\bf k})+\bar{\eta} U_1({\bf r},{\bf k})]e^{i {\bf k} \cdot {\bf r}}$ and $v_{{\bf k}}({\bf r}) =[V_0({\bf k})+\bar{\eta} V_1({\bf r},{\bf k})]e^{i {\bf k} \cdot {\bf r}}$.   The zeroth order quantum amplitudes are given by $U_0(k)=\sqrt{\frac{1}{2} \left(\frac{k^2+1}{E_B}+1 \right)}$ and $V_0(k)=\sqrt{\frac{1}{2} \left( \frac{k^2+1}{E_B}-1 \right)}$ and the first order Fourier transform of the quantum amplitudes are given by $\tilde U_1(\lambdav,{\bf k})$ and $\tilde V_1(\lambdav,{\bf k})$.

Below, we briefly sketch out the calculation for $F_{fluc}$, noting that the same analysis can be applied to $F_{cond}$, and then quote the results for both contributions to the force.  Since $\tilde V_1(\lambdav,{\bf k})$ is real, the only nonzero contribution to the force occurs when there is a pole on the real axis.  Separating the integral over the Fourier variable $\lambdav$ into Cartesian coordinates, we first perform the contour integration over $\lambda_z$.  For simplicity, we separate out the characteristic equation of the coupled Bogoliubov equations, $C(\lambdav,{\bf k})=-\lambda^4 - 2 \lambda^2 (1 + k^2 + 2 k_x \lambda_x) - 4 \lambda_x (k_x + k^2 k_x + k_x^2 \lambda_x + E_B q/\sqrt{2} - \lambda_x q^2/2)$, from $\tilde V_1(\lambdav,{\bf k}) = \Gamma_v(\lambdav,{\bf k})/C(\lambdav,{\bf k})$ where $\Gamma_v(\lambdav,{\bf k})$ can be easily deduced from the Bogoliubov-de Gennes equations.  The poles of $\Gamma_v(\lambdav,{\bf k})$ do not contribute to $F_{cond}$ because they are pure imaginary.  Of the four $\lambda_z$ poles of the characteristic equation, two of them, $z_{1,2}  = \pm \sqrt{-1-k^2-2 k_x \lambda_x - \lambda_x^2-\lambda_y^2 -\sqrt{1+E_B^2-2E_B \lambda_x+\lambda_x^2}}$, are pure imaginary for all values of parameter space and thus do not contribute to $F_{cond}$.  The other two, $z_{3,4} =\pm \sqrt{(\lambda_y^c)^2-\lambda_y^2}$ where $\lambda_y^c= \sqrt{-1-k^2-2 k_x \lambda_x - \lambda_x^2+\sqrt{1+E_B^2-2E_B \lambda_x+\lambda_x^2}}$, can be real in a certain region of parameter space, namely where both $\lambda_y^c$ is real and $-\lambda_y^c \le \lambda_y \le \lambda_y^c$.   We perform the contour integral over $\lambda_z$  to obtain
\begin{equation}
f_{fluc}({\bf k})=4 \pi   V_0 (k) \int_{-\infty}^{\infty}   d\lambda_x  \beta(\lambda_x,{\bf k}) I_{\Im(\lambda_y^c)=0}(\lambda_x)  \int_{-\lambda_y^c}^{\lambda_y^c} d \lambda_y \frac{1}{\sqrt{(\lambda_y^c)^2-\lambda_y^2}}
\end{equation}
where $I_{\Im(\lambda_y^c)=0}(\lambda_x)$ is the indicator function which is defined as unity when $\lambda_x$ is in subset $\Im(\lambda_y^c)=0$ and zero elsewhere; and $\beta(\lambda_x,{\bf k}) \equiv \lambda_x z_3 \hspace{0.5mm} \mathsf{Res}_{\lambda_z=-z_3 } \tilde V_1(\lambdav,{\bf k})$, $\mathsf{Res}_{x=x'}$ being the residue at $x=x'$ (note that $\beta(\lambda_x,{\bf k})$ is independent of $\lambda_y$ as $ \lambda_y$ cancels out).  The integral over $\lambda_y$ gives a constant $\pi$.  Integrating over {\bf k}, the integral over $\lambda_x$ in the parameter regime where $\lambda_y^c$ is real gives 
\begin{equation}
F_{fluc}=-32 \sqrt{2} \pi^{3/2} \bar{\eta}^2 p_0 \xi^2 \sqrt{n_0 a_{sc}^3} \int_0^\infty k^2 dk V_0(k) \left[ \int_{-1}^{f_k^c} d f_k \int_0^{\lambda_x^c} \beta(\lambda_x)+\int_{f_k^c}^1 d f_k \int_{-\lambda_x^c}^0 \beta(\lambda_x) \right]
\end{equation}
where $k_x=k f_k$, $f_k^c=-q E_B/\sqrt{2} k(k^2+1)$; and, solving the equation $\lambda_y^c=0$ for $\lambda_x$, we find $\lambda_x^c$ to be the only non-trivial real root.  At large $k$, the term in brackets goes as $k^{-1}$ and, since $V_0(k)$ goes as $k^{-2}$, the $k$-integral has a logarithmic ultraviolet divergence and we can write out the leading order term as 
\begin{equation}
F_{fluc} = \frac{16 \sqrt{2}}{3} \pi^{3/2} \bar{\eta}^2 p_0 \xi^2 \sqrt{n_0 a_{sc}^3} q  \ln \Lambda_{uv}
\end{equation}
where $\Lambda_{uv}$ is the dimensionless ultraviolet cutoff  \cite{kdep}.    Following a similar analysis, $F_{cond}=16 \sqrt{2} \pi^{3/2}\bar{\eta}^2 p_0 \xi^2 \sqrt{n_0 a_{sc}^3} q \ln \Lambda_{uv}$.  Thus, to leading order, the total force on an impurity in a moving, weakly interacting condensate at $T=0$ becomes 
\begin{equation}
F_{x} =\frac{64 \sqrt{2}}{3} \pi^{3/2} \bar{\eta}^2 p_0 \xi^2 \sqrt{n_0 a_{sc}^3} q \ln \Lambda_{uv} .
\end{equation}
(Recall $\bar{\eta} =2b/ a n_0 \xi^3$ is assumed to be small, $p_0=g n_0^2/2$, $\xi=(8 \pi n_0 a)^{-1/2}$, and $q$ is the Mach number, i.e. $q= c/c_s$.)
The ultraviolet divergence arises because of the singularity of the potential describing the point impurity.  However, if we use a potential with a nonzero width in the $x$-direction, the result becomes finite.  Therefore, in our problem, a sensible ultraviolet cutoff length is the dimensionless inverse of the scattering length describing the impurity-atom interaction $b$, i.e. $\Lambda_{uv} \approx \xi/b$.  
 It is also reassuring to note that in the limit that the scattering lengths tend to zero, the force disappears.  

It is important to emphasize that this force directly proportional to the flow speed, reminiscent of Stokes' drag, exists at all flow velocities, including those below Landau's critical velocity.  While we do expect this force to be dissipative - this can be proven by computing the time dependence of the condensate energy in the lab frame, for example - the existence of this force is a necessary but not sufficient condition for the system to be dissipative.  As discussed in \cite{roberts,roberts2}, this force is expected to eventually cancel out in a finite geometry due to backscattering effects.  However, for the point impurity considered in the present letter, the time scale (a unique signature of this effect) over which the force cancels out should be on the order of the volume of the system divided by the product of the scattering cross-section of the impurity and the speed of sound.   Moreover, as this force exists at all flow speeds in an infinitely extended geometry, one can speculate that the "persistence" of persistent currents arises from finite-size effects.

The author gratefully acknowledges many stimulating discussions with Yves Pomeau, Xavier Leyronas, and Masahito Ueda.  This work was financially supported by the Marie Curie Fellowship.


\begin{thebibliography}{9}
\bibitem{T0} Specifying the incoming $u_k(r)$ and $v_k(r)$ is sufficient to determine the scattered quasiparticle amplitudes if one assumes the normal physical boundary conditions, i.e. that there is causality and that there are no exponentially increasing scattered waves far from the point impurity.  Here, the incoming $u_k(r)$ and $v_k(r)$ are set by defining the spatial-temporal correlation functions (see \cite{roberts} for details) that fully determine the relevant quantum state, labeled "T=0".  Note that though the state at T=0 is not well-defined in this scattering problem, this label remains convenient because the state is always annihilated by $\hat\alpha$.
\bibitem{kdep} Note that at zero temperature the force on the impurity induced by the scattered fluctuations is dominated by the high-momentum fluctuations, whereas at low but nonzero temperatures the force is dominated by low-momentum phonons since, at such temperatures, phonon populations are concentrated at low momenta.
\end{thebibliography}
\end{document}